\documentstyle[12pt,epsfig]{article}
\newlength{\dinwidth}
\newlength{\dinmargin}
\begin{document}

%--------------------------------------------------------------------------
\def\bold#1{\setbox0=\hbox{$#1$}%
     \kern-.025em\copy0\kern-\wd0
     \kern.05em\copy0\kern-\wd0
     \kern-.025em\raise.0433em\box0 }
\def\slash#1{\setbox0=\hbox{$#1$}#1\hskip-\wd0\dimen0=5pt\advance
       \dimen0 by-\ht0\advance\dimen0 by\dp0\lower0.5\dimen0\hbox
         to\wd0{\hss\sl/\/\hss}}
\def\lq{\left [}
\def\rq{\right ]}
\def\II{{\cal I}}
\def\LL{{\cal L}}
\def\VV{{\cal V}}
\def\BB{{\cal B}}
\def\MM{{\cal M}}
\def\pv{\mbox{\bf p}}
\def\ovl{\overline}
\def\pr{^\prime}
%--------------------------------------------------------------------------
\newcommand{\be}{\begin{equation}}
\newcommand{\ee}{\end{equation}}
\newcommand{\bea}{\begin{eqnarray}}
\newcommand{\eea}{\end{eqnarray}}
\newcommand{\ba}{\begin{array}}
\newcommand{\ea}{\end{array}}
\newcommand{\nn}{\nonumber}
\newcommand{\dd}{\displaystyle}
\newcommand{\bra}[1]{\left\langle #1 \right|}
\newcommand{\ket}[1]{\left| #1 \right\rangle}
\newcommand{\spur}[1]{\not\! #1 \,}
\newcommand{\nor}[1]{{}_{\times}^{\times} #1 {}_{\times}^{\times}}
%--------------------------------------------------------------------------

\thispagestyle{empty}
\vspace*{1cm}
\rightline{Napoli DSF-NA-T-43/99}
\vspace*{2cm}
\begin{center}
  \begin{LARGE}
  \begin{bf}
A twisted conformal field theory description of the Quantum Hall
Effect
  \end{bf}
  \end{LARGE}

  \vspace{8mm}

  \begin{large}
Gerardo Cristofano ~~~~ Giuseppe Maiella \\ and \\ Vincenzo Marotta
  \end{large}
  \vspace{1cm}

\begin{it}
Dipartimento di Scienze Fisiche  \\
 Universit\'{a} di Napoli ``Federico II'' \\
and \\
INFN, Sezione di Napoli
\end{it}
\end{center}
\begin{quotation}

\begin{center}
\begin{bf}
Abstract\\
\end{bf}\end{center}
We construct an effective conformal field theory by using a procedure which
induces twisted boundary conditions for the fundamental scalar fields. That
allows to describe a quantum Hall fluid at Jain hierarchical filling,
$\nu=\frac{m}{2pm+1}$, in terms of one charged scalar field and $m-1$
neutral ones. Then the resulting algebra of the chiral primary fields is
$U(1){\times} {\cal W}_m$. Finally the ground state wave functions are given as
correlators of appropriate composite fields ($a$-electrons).

\noindent

\vspace*{0.5cm}


%PACS: 11.25.Hf, 02.20.Sv, 03.65.Fd

\vfill
{\small\bf
\begin{tabbing}
Postal address: Mostra d'Oltremare Pad.19-I-80125 Napoli, Italy \\ E:mail:
\=  gerardo.cristofano(giuseppe.maiella;vincenzo.marotta)@na.infn.it
\end{tabbing}}
\end{quotation}

\newpage
\baselineskip=18pt
\setcounter{page}{2}

In this letter we investigate the possibility of constructing a
Conformal Field Theory (CFT) description of the plateaus in the
Jain hierarchical model with filling fractions
$\nu=\frac{m}{2pm+1}$ \cite{jain}. In our proposal the level
$l=\{1,\dots , m\}$ of the hierarchy is generated from a CFT
description \cite{cft} of $\nu=1$ Quantum Hall Effect (QHE) by an
induction procedure. Furthermore an unified Effective Conformal
Field Theory (ECFT) framework is given in which the relevant edge
states and gapless excitations for any quantum Hall plateau (both
of the Integer Quantum Hall Effect (IQHE) and Fractional Quantum
Hall Effect (FQHE)) are naturally described.

Historically, the motivations for the use of ECFT in describing
the QHE at the plateau go back to the observation that the ground
state wave function (Laughlin states) for the filling
$\nu=1/(2p+1)$ \cite{laugh} can be written as correlator of
vertex operators describing the anyon states (charged states).
Such operators were identified as primary states of a 2D chiral
CFT with central charge $c=1$ \cite{lufub,cgm2}. For general
filling it is known that it is necessary to introduce neutral
fields in order to describe the area-preserving edge deformations
of the incompressible fluid. In this context two classes of CFTs
have been proposed for Jain plateaus: the multi-component bosonic
theory \cite{frohlich}, characterised by the symmetry
$\widehat{U(1)}{\times}\widehat{SU(m)}_{1}$ and the $W_{1+\infty}$
minimal models \cite{ctz5} both with central charge $c=m$. In
spite of having the same spectrum of edge excitations, they
manifest differences in the degeneracy of the states and in the
quantum statistics.

Our aim is to give an alternative and very simple construction of the
hierarchical model based upon well-known techniques of CFT. The resulting
theory is related to an orbifold construction of the above CFTs. Only one
$U(1)$ charged current survives to the discrete twist group induced by our
procedure and the resulting algebra $\widehat{U(1)}{\times} {\cal W}_m$ can be
viewed as a RCFT extension of the minimal models \cite{KT}.

In this letter we first briefly present the physical arguments underlying
the ECFT description of the QHE and we summarize the two main proposals for
including in this scheme the hierarchical model.

We then present a version of the $m$-reduction procedure
introduced in \cite{FKN} in the framework of 2D quantum gravity
which allows us to obtain a $c=m$ (daughter) CFT from a $c=1$
(mother) CFT.The daughter theory preserves the $W_{1+\infty}$
symmetry of the mother theory, which implies the
incompressibility of Quantum Hall Fluid (QHF) at the plateaus. In
fact it is well known that for any positive integer $m$,
$W_{1+\infty}$ with central charge $c$ is isomorphic to its
subalgebra consisting of elements of degrees divisible by $m$ and
which appear with central charge $m c$ \cite{FKN, kac}.

Further we discuss in detail how starting from one scalar (chiral) field
the $m$-reduction procedure generates the image of $m$ scalars which can be
divided in one charged component with non zero momentum and $m-1$ neutral
ones without zero modes because of the induced twisted boundary conditions.
Then it is shown that the full symmetry of this theory is $U(1){\times}{\cal
W}_{m}$.

Finally we present the construction of the ground state wave functions for
the Jain hierarchy scheme at $\nu=\frac{m}{2pm+1}$. It is shown that the
primary fields which are composite operators have the correct monodromy
property and chirality. This is due to a cancellation between the
non-analytic behaviour of the charged component and that of the neutral
one.

The ECFTs description of QHE at the plateau is justified by the
incompressibility of the Laughlin fluid. The dynamical symmetry
of this fluid in the disk geometry is the area-preserving
diffeomorphisms of the plane which implies the $W_{1+\infty}$
algebra. This is the unique centrally extended quantum analogue
of the classical area preserving diffeomorphisms algebra
$w_{\infty}$ on the circle.

The infinite generators $W^{n+1}_m$ of $W_{1+\infty}$ of conformal spin
($n+1$) are characterized by a mode index $m\in Z$ and satisfy the algebra:
\be
\left[W^{n+1}_m,W^{n'+1}_{m'}\right]=(n'm-n m')W^{n+n'}_{m+m'}+
q(n,n',m,m')W^{n+n'-2}_{m+m'}+
d(n,m)c \,\delta^{n,n'}\delta_{m+m'=0}
\ee
where the structure constants $q$ and $d$ are polynomials of
their arguments, $c$ is the central charge, and dots denote a
finite number of similar terms involving the operators
$W^{n+n'-2l}_{m+m'}$ \cite{BS, ctz}.

Such an algebra contains an Abelian current $\widehat{U}(1)$ for $n=0$ and
a Virasoro algebra for $n=1$ with central charge $c$. Their zero modes
eigenvalues are identified as the charge and angular-momentum of the edge
excitations in the QHF.

A large class of rational extensions (RCFTs) of $W_{1+\infty}$
with the same Virasoro algebra exists. But one can construct a
RCFT which has $\widehat{U}(1){\times}\widehat{SU}(m)_1$ as extended
symmetry for the filling $\nu=\frac{m}{2mp+1}$. Such a theory in
particular contains $m$ independent Abelian currents
\cite{frohlich}. On the other hand a description of Jain plateaus
has been given \cite{ctz5} in terms of a $c=m$ CFT with full
degenerate $W_{1+\infty}$ representations. They are isomorphic to
$\widehat{U}(1){\times}{\cal W}_m$ theory, where ${\cal W}_m$ is the
$p\rightarrow
\infty$ limit of the Zamolodchikov-Fateev-Lykyanov models with
$c=(m-1)\left(1-\frac{n(n+1)}{p(p+1)}\right)$ \cite{fateev}.

The main differences with respect to the
$\widehat{U}(1){\times}\widehat{SU}(m)_1$ theory are the following: 1)
There is only one Abelian current instead of $m$ independent
ones. 2) There are neutral excitations that cannot be associated
to $m-1$ independent edges. 3) The multiplicity of the states is
different: In particular minimal models have only one state for
any ${\cal W}_m$ irreducible representation.

Our approach is to describe all the stable plateaus starting from
the filling $\nu=1$ in terms of a constrained boson, for which
the relevant ECFT is a $c=1$ bosonized free Dirac fermion
described by a scalar chiral field compactified on a circle with
radius $R^2=1$. Then the $U(1)$ current is given by
$J(z|1,1)=i\partial_z Q(z)$, where $Q(z)$ is the compactified
Fubini-Veneziano field with the standard mode expansion:
\be
Q(z)=q-i\, p\, ln z + \sum_{n\neq 0}\frac{a_n}{n}z^{-n}
\ee
where $a_n$, $q$ and $p$ satisfy the commutation relations $
\left[a_n,a_{n'}\right]=n\delta_{n,n'}$ and $\left[q,p\right]=i $.

The representations are realized by the vertex operators $
U^{\alpha}(z)=:e^{i\alpha Q(z)}: $ with $\alpha^2=1$ and
conformal dimension $h=\frac{1}{2}$. Furthermore, the theory
contains the Virasoro algebra, with central charge $c=1 $,
generated by the stress-energy tensor
$T(z|1,1)=-\frac{1}{2}:\left(\partial_z Q(z)\right)^2:$. The
chiral algebra can be extended to the full $W_{1+\infty}$ with
central charge $c=1$ as it is well known giving the $W^{n+1}(z)$
fields in terms of the product of the current $J(z|1,1)$ and
their derivatives \cite{ctz5}.

To obtain the holomorphic part of the ground state wave function we need to
consider the correlator:
\be
<N_e\alpha|\prod_{i=1}^{N_e}U^{\alpha}(z_i)|0>=
\prod_{i<j}^{N_e}(z_i-z_j)
\ee
where the momentum $N_e\alpha$ assures the neutrality condition
of the fluid \cite{lufub,cgm2}.

In order to construct the $\nu=m$ filling we start with the set
of fields in the above CFT (mother). By using the $m$-reduction
procedure, which consists in considering the subalgebra generated
only by the modes which are a multiple of an integer $m$, we get
the image of an orbifold of a $c=m$ CFT. This kind of subalgebras
have been studied in a few cases \cite{FKN,KW,B,VM,BHS}.

The fields in the mother CFT can be factorized into irreducible orbits of
the discrete group $Z_m$ which is a symmetry of the daughter theory. Thus
we split these fields into components which have well defined
transformation properties under this group.

In order to compare the image so obtained to the $c=m$ CFT, we  map
$z\rightarrow z^{1/m}$ and we will indicate the components in the base
$\hat{z}=z^{m}$ with an hatted symbol (for instance, $\phi(z)\rightarrow
\widehat{\phi}(z)$). In particular, any component in the subalgebra is a
function only of the variable $z^m$. The above conformal map
needs to be taken with care because only the reduced Virasoro
algebra contains the correct generators of this transformations
\cite{VM}.

Therefore we introduce the invariant scalar field
\be
X(z|m,1)=\frac{1}{m}\sum^{m}_{j=1}Q(\varepsilon^j z)
\ee
where $\varepsilon^j=e^{i\frac{2\pi j}{m}} $, corresponding to a
compactified boson on the circle but with radius $R_X^2= 1/m$. This field
depends only by powers of $z^m$ and modes $a_{nm}$ and satisfies trivial
boundary conditions. It is the basic field of the $U(1)$ electric charged
sector of the theory.

The not invariant part of $Q(z) $ can be organised in the image of $m$
constrained bosons
\be
\phi^j(z|m,1)=Q(\varepsilon^j z)-X(z|m,1)
\ee
with the condition $\sum_{j=1}^{m}\phi^j(z|m,1)=0$.

These fields satisfy non-trivial twisted boundary conditions
\be
\alpha{\cdot}\phi^j(\varepsilon z|m,1)=\alpha{\cdot}\phi^{j+1}(z|m,1)+2\pi n \alpha{\cdot}p
\,\,\, n\in Z \label{eq: shift}
\ee
where the shift is due to the definition of index $j$ mod $m$.

In Ref.\cite{VM}, it was also defined an isomorphism between fields on the
$z$ complex plane and fields on the $z^m$ plane by means of the following
identifications:
\be
a_{nm+l} \longrightarrow \sqrt{m}a_{n+l/m} \hspace{1cm} q
\longrightarrow \frac{1}{\sqrt{m}}q  \label {eq: 32}
\ee

The $J(z|1,1)$ current of the mother theory decompose into the a charged
current given by $J(z|m,1)=\partial_z X(z|m,1)$ and $m-1$ neutral ones
$\partial_z\phi^j(z|m,1)$. It is useful to notice that the neutral currents
can be expressed in a new basis in which the boundary conditions are
diagonal:
\be
\partial_z\phi_l(z|m,1)=\sum_{j=0}^{m-1}\varepsilon^{-j l}
\partial_z\phi^j(z|m,1)
\ee
and by using eq.(\ref{eq: 32}) it is simple to show that they
have a Laurent expansion in terms of new modes given by:
\be
\widehat{\partial_z\phi}_l(z|m,1)=\sum_{n\neq 0}a_{n-l/m} z^{-n-1+l/m}
\ee

This property is typical of currents in the twisted sector induced by a
twist field placed at the origin.

In the same way, every vertex operator in the  mother theory can be
factorized in a vertex that depends only on the invariant field:
\be
{\cal U}^{\alpha}(z|m,1)=z^{\frac{\alpha^2}{2}\frac{(m-1)}{m}}:e^{i\alpha{\cdot}
X(z|m,1)}: ~~~\alpha^2=1
\ee
and in vertex operators depending on the $\phi^j(z|m,1)$ fields. We
introduce the neutral components:
\be
\psi^{\alpha}_1(z|m,1)=\frac{z^{\frac{\alpha^2}{2}\frac{(1-m)}{m}}}{m}\sum_{j=1}^{m}
\varepsilon^{\frac{\alpha^2 j}{2}} :e^{i\alpha{\cdot} \phi^{j}(z|m,1)}:
\ee
which satisfy the fundamental product:
\bea
&&\psi_1^{\alpha}(z|m,1)\psi_1^{\beta}(\xi |m,1)
\nn \\ &&=
\frac{z^{\frac{\alpha^2}{2}\frac{1-m}{m}}
\xi^{\frac{\beta^2}{2}\frac{1-m}{m}}}{m^2}
\sum_{j,j'=1}^{m}\varepsilon^{\frac{\alpha^2 j+\beta^2 j'}{2}}
:e^{i\alpha{\cdot}\phi^{j'}(z|m,1)}
e^{i\beta{\cdot}\phi^j(\xi |m,1)}:
\frac{(\varepsilon^{j'}z-\varepsilon^{j}\xi)^{\alpha
{\cdot} \beta}}{(z^m-\xi^m)^ {\frac{\alpha {\cdot} \beta}{m}}}
\label {eq: 31}
\eea
and have conformal dimensions given by eq.(\ref {eq: CD}).

The set of primary fields generated by this product can be given in terms
of the fundamental representations $\Lambda^i$ of $SU(m)$ Lie algebra. In
fact, defining
\be
\phi^{\Lambda^i}(z|m,1)=\sum_{j=1}^{i}\phi^j(z|m,1)
\ee
and $\phi^{\Lambda}(z|m,1)$, where
$\Lambda=\sum_{i=1}^{m-1}l_i\Lambda_i$ and introducing the
$m$-ality parameter $a=\sum_{i=1}^{m-1}i l_i$ $(mod \, m) $,
which is invariant under the addition of any vector in the root
lattice, the exact form of these fields can be deduced by the
analysis of the OPE of eq.(\ref{eq: 31}) for $\alpha=\beta$ to
get the $a=2$ field. By repeated application of this analysis we
can obtain the full set of fields:
\be
\widehat{\psi}^{\alpha}_a(z|m,1)=
\sum_{j_1>j_2>\dots >j_a}f(\varepsilon^{j_1},\dots,\varepsilon^{j_a},z^{1/m})
:e^{i\alpha\widehat{\phi}^{j_1}(z|m,1)}\dots e^{i\alpha\widehat{\phi}^{j_k}
(z|m,1)}:
\ee
where the functions $f(\varepsilon^{j_1},\dots,\varepsilon^{j_a},z)$ can be
extracted from the OPE relations. The sum takes into account the fact that
any field can be associated to the $a$-th fundamental representation of
$SU(m)$ (namely, the antisymmetric tensor representation) and their
operator algebra is
\be
\widehat{\psi}_{a} ^{\alpha}(z|m,1) \widehat{\psi}_{a'} ^{\alpha}(\xi |m,1)
=\frac{C_{a,a'}}{(z-\xi)^{\frac{a a'}{m}}}
\left[\widehat{\psi}_{a+a'} ^{\alpha}(z|m,1)+O(z-\xi)\right] \,\,\, a+a'<m
\ee
while for the $a+a'=m$ case it contains ${\cal W}_m$ generators:
\be
\widehat{\psi}_{a} ^{\alpha}(z|m,1) \widehat{\psi}_{m-a} ^{\alpha}(\xi |m,1)
=\frac{C_{a,a'}}{(z-\xi)^{\frac{a (m-a)}{m}}}
\left[1+ \widehat{T}(z|m,1)(z-\xi)^2+O(z-\xi)^3\right]
\ee

Moreover, the $SU(m)$ representations that can appear are the fundamental
ones $\Lambda_a$ because the OPE algebra of $\psi^{\alpha}_1(z|m,1)\equiv
\psi^{\alpha}_{\Lambda_1}(z|m,1)$ gives only fields up to
$\psi^{\alpha}_{\Lambda_{m-1}}(z|m,1)$ while
$\psi^{\alpha}_{\Lambda_{m}}(z|m,1)$ is the identity operator.

Notice that no neutral currents are present in the above OPE, as one would
expect from symmetry considerations.

It is well known that the $c_X=1$ RCFT with $R_X^2=1/m$ has $m$
primary fields $\widehat{{\cal U}}^{\frac{a}{\sqrt{m}}}(z|m,1)$
that can be parametrized by $\alpha=\frac{a}{\sqrt{m}}$
$a=1,\dots ,m$ and have conformal dimensions
$h_a=\frac{a^2}{2m}$. In our formalism, these fields appear
together with the neutral ones giving rise to the $m$ fundamental
composite operators:
\be
\widehat{V}^{\frac{a}{\sqrt{m}}}(z|m,1)=\widehat{{\cal
U}}^{\frac{a}{\sqrt{m}}}(z|m,1)\widehat{\psi}_a(z|m,1)
\label {eq: ANI}
\ee

From these primary fields we can obtain the new $W_{1+\infty}$ with central
charge $c=m$ which contains in particular the spin two operator
$\hat{T}(z|m,1)$, which is the generator of the Virasoro algebra. It is the
sum  of two independent operators, one depending on the charged sector:
\be
\widehat{T}_{X}(z|m,1)=-\frac{1}{2} :\left(\widehat{\partial_{z}X}(z|m,1)
\right)^2:
\label {eq: VIR1}
\ee
and the other given in terms of the $Z_m$ twisted bosons
$\widehat{\phi}^j(z|m,1)$:
\be
\widehat{T}_\phi(z|m,1)=-\frac{1}{2}\sum_{j,j'=1}^{m}
:\widehat{\partial_z\phi}^j(z|m,1)\widehat{\partial_z\phi}^{j'}(z|m,1):+\
\frac{m^2-1}{24 m z^2} \label {eq: VIR2}
\ee

It is not very hard to verify that the conformal dimensions of the $m$
fields of eq.(\ref{eq: ANI}) are:
\be
h_a=\frac{a^2}{2m}+\frac{a}{2}\left(\frac{m-a}{m}\right)=\frac{a}{2}
\,\,\,\, a\in\{1,\dots,m\} \label {eq: CD}
\ee
while their electric charge is given by the eigenvalue of the $U(1)$
current and is always an integer:
\be
Q_a=a
\ee

Higher spin currents in $W_{1+\infty}$ algebra are given by the infinite
generators in the enveloping algebra of the $U(1)$ charged sector and by
the ${\cal W}_m$ currents obtained from the neutral sector. The explicit
form for $n\leq 4$ was given in \cite{FKN}. We report here only the first
element beyond the spin two of the series which in our basis is expressed
as:
\be
\widehat{W}^3(z|m,1)=\frac{1}{2\sqrt{m}}\sum_{j,j',j''=1}^{m}
:\widehat{\partial_z\phi}^j(z|m,1)\widehat{\partial_z\phi}^{j'}(z|m,1)
\widehat{\partial_z\phi}^{j''}(z|m,1):
\ee

Now we study the Jain filling fractions $\nu=\frac{m}{2pm+1}$ which
naturally arise in our approach. In fact the appropriate CFT is obtained
with a $2pm$ flux attachment starting from the $\nu=m$ filling.

In order to do so, we factorize the fields into two parts, the first is the
$c_X=1$ charged sector with radius $R_X^2=\frac{2pm+1}{m}$, the second
describes neutral excitations with total conformal central charge
$c_{\phi}=m-1$ for any $p\in N$.

The $U(1)$ sector is now described by the compactified boson $X(z|m,2pm+1)$
and its related vertex operators $\widehat{{\cal
U}}^{{\pm}\alpha_l}(z|m,2pm+1)$, with $\alpha_l=l/\sqrt{m(2pm+1)}$, $l=1,\dots
, m (2pm+1)$, which produce excitations with anyonic statistics
$\theta=\pi\alpha_l^2$. While the $m-1$ neutral bosons $\phi^j(z|m,1)$ are
independent from the flux number $p$ and satisfy Abelian $Z_m$ boundary
conditions.

To obtain a pure holomorphic function we will consider the correlator of
the composite operators ${\cal V}^{\alpha_l}(z|m,2pm+1)={\cal
U}^{\alpha_l}\psi^{\alpha}_{a}(z|m,2pm+1)$ with conformal dimension:
\be
h_l(2pm+1)=\frac{l^2}{2m(2pm+1)}+
\frac{a}{2}\left(\frac{m-a}{m}\right); ~~~~~~~~l=1,2,\dots ,m(2pm+1)
\ee
and electric charge $Q_l=\frac{l}{2pm+1}$.

Notice that there are integer charge quasi-particles (from now on to be
referred to as $a$-electrons) which have semi-integer (or integer)
conformal dimension given by:
\be
h_l(2pm+1)=a^2p+\frac{a}{2} ~~~l=(2pm+1)a; ~~~~~~~~~~a=1,2,\dots ,m
\ee

This follows by the construction of the Virasoro algebra with
central charge $c=m$ as it has been done for the case $\nu=m$
(see eq.(\ref{eq: VIR1},\ref{eq: VIR2})). We should point out
that $m$-ality in the neutral sector is coupled to the charged
one exactly as it was derived in \cite{frohlich,ctz5} by physical
request of the locality of electrons with respect to all the edge
excitations. This follows from the fact that our projection when
applied to a local field (namely the electron field for $\nu=1$),
automatically couples the discrete $Z_m$ charge of $U(1)$ with
the neutral sector, in order to give a totally single-valued
composite field.

Also notice that the $m$-electron vertex operator does not contain any
neutral field. Therefore the $m$-electron wave function is realized only by
means of the $c_X=1$ charged sector, so that is a pseudoparticle with
electric charge $m$ and magnetic charge $2pm+1$ \cite{cgm1}.

We are now ready to give the holomorphic part of the ground state
wave function for the Jain filling $\nu=\frac{m}{2pm+1}$. Then we
consider the $N_e$ single($a=1$)-electrons correlator which
factorizes into a Laughlin-Jastrow type term coming from the
charged sector:
\be
<N_e\alpha|\prod_{i=1}^{N_e}\widehat{{\cal U}}^{\alpha}(z_i|m,2pm+1)|0>
=\prod_{i<i'=1}^{N_e}(z_i-z_{i'})^{2p+\frac{1}{m}}
\ee
and a contribution coming from the neutral excitations:
\be
<0|\prod_{i=1}^{N_e}\widehat{\psi}_1
^{\alpha}(z_i|m,1)|0>
=\frac{\sum_{\{j_{i}\}=1}^{m}\varepsilon^{\frac{\alpha^2}{2}(2i-1) j_i+ j_{i}}
\prod_{\{j_{i},j_{i'}\}=1}^{m}
(\varepsilon^{j_{i}}z^{1/m}_i-\varepsilon^{j_{i'}}
z^{1/m}_{i'})^{\alpha^2}}{\prod_{i<i'=1}^{N_e}(z_{i}-z_{i'})^{\frac{1}{m}}}
\label{eq: FUN}
\ee

We observe that the non analytic part of the neutral fields $\widehat{\psi}
^{\alpha}(z_i|m,1)$ is necessary to eliminate the non integer part of the
exponent in the correlator of the charged  fields. In particular the  four
point function for the fillings $\nu=\frac{2}{4p+1}$ is given by
\bea
<4\sqrt{\frac{4p+1}{2}}|\prod_{i=1}^{4}\widehat{V}^{\sqrt{\frac{4p+1}{2}}}
(z_i|m,2pm+1)|0>&=&-\frac{1}{8}
\prod_{i<i'=1}^{4}(z_{i}-z_{i'})^{2p}{\times}  \nn \\
\left\{\left(\sqrt{z_1 z_3}+\sqrt{z_2 z_4}\right)(z_1-z_3)(z_2-z_4)+ ~~
permutations \right\}
\eea

Even though it is hard to work out explicitly the sum over the
phases in eq.(\ref{eq: FUN}) for the general $N_e$ point
functions we found some general rules:

a) Any non null wave function can be written as cluster of $m$
one-electrons fields,

b) there are no zero's for particles belonging to the same cluster,

c) the zero's for particles in different clusters are of order one.

We must notice that these clustering properties are found in the context of
the pairing phenomena in the QHE \cite{MR}. In fact our methods can be
extended to the paired Hall states (i.e. $\nu=\frac{m}{pm+2}$) allowing for
a description of filling $\nu=\frac{m}{pm+2}$ \cite{CGM}. In this last case
the neutral modes describe parafermions and contribute to the ground state
wave function with a generalized Pfaffian term.

In conclusion our $m$-reduction procedure allows for a chiral
projection, to the lowest Landau level, which gives rise to a
local description of the ground state through the cuts
cancellation, so that the incompressibility of the QHF is
preserved.

\bigskip
{\bf Acknowledgments} - We thank A. Cappelli, G. Zemba  and A. Sciarrino
for useful comments and for reading the manuscript.

\end{document}